\documentclass[preprintnumbers,amsmath,amssymb]{revtex4}
\usepackage{graphics,graphicx}
\usepackage{dcolumn}
\usepackage{subfigure}
\makeatletter
\begin{document}

\title{City Size Distributions For India and China}
\author{Kausik Gangopadhyay}
\email{kausik@iimk.ac.in} \affiliation{Indian Institute of
Management Kozhikode, IIMK Campus P.O., Kozhikode 673570, India}
\author{B. Basu}
\email{banasri@isical.ac.in, Fax:91+(033)2577-3026}
\affiliation{Physics and Applied Mathematics Unit\\
 Indian Statistical Institute\\
 Kolkata-700108, India }
%
%
\begin{abstract}
\begin{center}
{\bf Abstract}
\end{center}
This paper studies the size distributions of urban agglomerations for
India and China. We have estimated the scaling exponent for the
Zipf's law with  the Indian census data for the years of 1981-2001
and the Chinese census data for 1990 and 2000. Along with  the
biased linear fit estimate, the maximum likelihood estimate for the
Pareto and Tsallis $q$-exponential distribution has been computed.
For India, the scaling exponent is in the range of
 [1.88, 2.06] and for China, it is in the interval [1.82, 2.29].
 The goodness-of-fit tests of the estimated distributions are performed using the
 Kolmogorov-Smirnov statistic.
\end{abstract}
pacs: 89.75.Da, 89.65.-s, 89.65.Gh

\maketitle
\section{Introduction}
Nature, in spite of its complex character, often displays
macroscopic regularity, which can be described broadly by simple
laws at different scales. Examples include the size distribution of
islands \cite{land} and lunar craters \cite{lun}, occurrence of
forest fires \cite{forest} and solar flares \cite{solar},
websurfings \cite{web}, wealth and income distribution in societies
\cite{bkc},  and also football goal distribution \cite{football}.
However, the size-distributions of populations in cities aces them
all in arresting our longest attention  \cite{chris}.

It is conjectured that the city sizes obey a surprisingly
 simple  law, known as Zipf's law \cite{zipf} (alternatively known as  Pareto
 distribution or simply power law),
 which states that
  the population-wise rank, $R_x$, of a city with $x$ number of inhabitants
  is proportional to $x^{-\alpha}$, with $\alpha$ being close to one.
  This law has received empirical endorsement from
  different studies using data from USA, Switzerland \cite{switzerland}, Brazil \cite{pro1} etc.
However, fewer analysis have been conducted for urban agglomerations with comparatively
lower populations. For example, a recent work \cite{jap} shows that the size distribution of towns and villages are completely different from that of cities. Another interesting  study \cite{stanley1,stanley2} looks at the spatial distribution of city population
and observes deviation from the usual power law. Some other studies reveal that
a deviation of the power law may be observed if all the urban
agglomerations of an urbanized nation is considered
\cite{japan,mend}. Some alternative statistical distributions
\cite{mandl,weibl,mend} are suggested to include the deviation from
the power law.
In general, the distribution of urban agglomerations in USA can be
well-described by using a Tsallis $q$-exponential distribution
\cite{mend}, which is an extension of the standard Zipf-Mandelbrot
law \cite{mandl} proposed in the context of generalized statistical
mechanics \cite{tsal}.

The empirical phenomenon of the power law receives the theoretical
support from some mathematical models \cite{mars,simon,dover}. These
theories model the evolution of the distribution of the city-size
either as a time dependent process, or as a result of interactions
among individuals.  In both cases, reality is an approximation of
the theoretical prediction of the limiting distribution and the
empirically observed distribution converges to its limiting value
depending on either the time-length of the process, or the total
population of the considered society. This motivates us to analyze
empirically the population distribution of cities for the two most
populous countries of the world such as India and China, which are
arguably the foremost ancient civilizations as well. These two
countries are comparatively less urbanized  and possess remarkably
heterogeneous socio-economic structures. In any case, the data from
India and China should be ideal to test the theoretical limiting
predictions.

In this work, we carry out an empirical investigation with census
data of India  for the time period of 1981-2001 as well as with the
Chinese census data for the years of 1990 and 2000. We estimate the
scaling exponents for the Pareto distribution and also for a more
generalized Tsallis $q$-exponential distribution. Also, goodness-of-fit tests for these hypothesized distribution have been
performed. The methodology is described in Section
\ref{methodology}; whereas the data and the results are discussed in
Section \ref{section_results}. We conclude our study in the final
section.



 \section{Methodology}
\label{methodology}
 Let $p(\cdot)$ be a probability density function of the city-size distribution. The corresponding cumulative distribution
  function (CDF) and the complementary cumulative distribution function (CCDF) are given by $P(\cdot)$ and $P^C(\cdot)$,
  respectively. The CDF, $P(x)$ is the probability that a city has a population less than or equal to x and the CCDF, $P^C(x)$, is the  probability that a
city has a population greater  than x.
  By definition,
\begin{equation}
 P(x)=\int_0^x
p(x^\prime) dx^{\prime};~~~~P^C(x) = 1 - P(x)
\end{equation}
In  case of city-size distribution following the Zipf's law,
\begin{equation}
p_{\alpha}(x) = Cx^{-\alpha}~~~ \mbox{and}~~~ \displaystyle
P_{\alpha}^C(x)=\frac{C}{\alpha-1} x^{-(\alpha-1)} \label{Pareto}
\end{equation}
where $\alpha$ and $C$ are constants. $\alpha$ is called the
exponent of the power law. This family of power law distributions
for $\alpha
>1$ are known as the Pareto distribution. From equation (\ref{Pareto}),
it is obvious that $p_{\alpha}(x)$ diverges to infinity for any
value of $\alpha
> 1$ as $x \rightarrow 0$. Therefore, some minimum value, $x_{min}$,
is usually considered for the support of the Pareto distribution.
The corresponding probability density function, the CDF and the CCDF
are given by:
\begin{equation}
p_{\alpha}(x) = \frac{(\alpha-1)\cdot
x_{min}^{\alpha-1}}{x^{\alpha}};~~~~~~~P_{\alpha}(x) = 1 -
\left(\frac{x_{min}}{x}\right)^{\alpha-1};~~~~~~~P_{\alpha}^C(x) =
\left(\frac{x_{min}}{x}\right)^{\alpha-1} .\label{Pareto_dist}
\end{equation}

A more general distribution, namely the Tsallis $q$-exponential
distribution,  has been proposed in \cite{mend}. The probability
density function, the CDF and the CCDF of this distribution, as
given in \cite{shalizi_MLE}, are noted below:
\begin{equation}
 p_{\theta, \sigma} (x) =
\frac{\theta}{\sigma}\left(1+\frac{x}{\sigma}\right)^{-\theta-1};~~~~~
 P_{\theta, \sigma} (x) =
1- \left(1+\frac{x}{\sigma}\right)^{-\theta} ;~~~~~
 P_{\theta, \sigma}^C (x) =
\left(1+\frac{x}{\sigma}\right)^{-\theta}\label{Tsallis_dist}
\end{equation}

From the equations (\ref{Pareto_dist}) and (\ref{Tsallis_dist}), it
is evident that the two distributions of Pareto and Tsallis
q-Exponential are approximately identical for large values of $x$,
when we set $\theta$ to $(\alpha -1)$ and $\sigma$ to $x_{min}$.

The slope of the plot, in which log of the rank of a city,
$\log(R_x)$, is plotted against the log of its population,
$\log(x)$, has been used to estimate the exponent of the power law
in almost all the previous studies. It has been shown
\cite{shalizi_powerlaw} that this produces a biased estimate of the
power law exponent. Alternatively the Maximum Likelihood Estimator
(MLE) produces the most efficient estimate. For a sample consisting
cities with populations $x_1$, $x_2$, ..., $x_n$, the log-likelihood
of the sample is described by the following expression:
\begin{equation}
l(\underline{x}; \nu) = \sum_{i=1}^{n} \log(p(x_i; \nu))
\label{log-likelihood}
\end{equation}
where $p(x_i; \nu)$ is the probability density function of the
$i^{th}$ observation $x_i$ drawn from a certain distribution with
parameter $\nu$. The function $l(\underline{x}; \nu)$ is maximized
with respect to the parameter $\nu$ to derive the maximum likelihood
estimate $\widehat{\nu}$ of the parameter $\nu$. Mathematically,
\begin{equation}
\widehat{\nu} = \underset{\nu}{\arg \max} ~~~l(\underline{x}; \nu).
\label{MLE}
\end{equation}
In particular, the MLE for the Pareto distribution with $x_{min}$ as
the minimum value is given by:
\begin{equation}
\widehat{\alpha}_{MLE} = 1+ n \left[\sum_{i=1}^n
\log\left(\frac{x_i}{x_{min}}\right)\right]^{-1}
\end{equation}
The solution of the following system of simultaneous equations
\cite{shalizi_MLE} represents the MLE for the Tsallis
$q$-exponential distribution with parameters $(\theta, \sigma)$ and
$x_{min}$ as the minimum value:
\begin{equation}
\displaystyle
\begin{array}{c}
\widehat{\theta}_{MLE} =  n \left[\sum_{i=1}^n
\log\left(\frac{1+x_i/\widehat{\sigma}_{MLE}}{1+x_{min}/\widehat{\sigma}_{MLE}}\right)\right]^{-1}
\\
\widehat{\sigma}_{MLE} = -\widehat{\theta}_{MLE}
\frac{x_{min}}{1+x_{min}/\widehat{\sigma}}_{MLE} +
\frac{\widehat{\theta}_{MLE}+1}{n} \sum_{i=1}^{n}
\frac{x_i}{1+x_i/\widehat{\sigma}_{MLE}}
\end{array}
\end{equation}
The Fisher's Information matrix \cite{crrao} gives the asymptotic
variance of the MLE. We can also compute the standard error in our
estimate by the technique of Bootstrapping \cite{boot}. In this
method, we draw sub-samples from our original sample and compute the
maximum likelihood estimates for those sub-samples. The standard
error in the estimates obtained from different sub-samples is our
estimate for the standard error of the MLE.

So far, we have treated $x_{min}$ as an exogenous parameter in
estimating the scaling exponent. To derive \cite{shalizi_powerlaw}
an endogenous value for $x_{min}$,  it is required to minimize
 the distance between the two CDFs,  one obtained from
the data and the other arising out of the best-fitted power law
model, contingent on the value of $x_{min}$. In general, if we
choose $\hat{x}_{min}$ higher than the true value of $x_{min}$, then
the size of the data set is effectively reduced. Due to statistical
fluctuations, a reduced data-set augments the error level for the
empirical distribution, when compared with the fitted theoretical
distribution. On the other hand, if $\hat{x}_{min}$ is smaller than
the true value of $x_{min}$, the distribution will differ because of
the fundamental difference between the data and the fitted model.
Kolmogorov-Smirnov (KS) statistic \cite{KS} is a standard measure to
quantify this distance, $D$, between the two probability
distributions with CDFs $F_1(\cdot)$ and $F_2(\cdot)$.
Mathematically speaking,
\begin{equation}
D  = \underset{x}\sup |F_1(x) - F_2(x)|
\end{equation}
It may be worth noting that the CDF of the best-fitted
power law depends on the choice of $x_{min}$ and  $D$ is minimized
with respect to this $x_{min}$. This leads to the optimal model,
which is the closest one to the empirical distribution among the
class of best fitted models. Simultaneously, we obtain $\hat{x}_{min}$, the optimal
estimate for $x_{min}$.

\subsubsection*{Goodness-of-fit Tests}
It might be interesting to test the null hypothesis \cite{crrao}  of
empirical distribution following our estimated distribution (Pareto
or Tsallis $q$-exponential).  It should be mentioned here that even
if we estimate the parameters of our distribution using the
empirical observations, it is not anyway imperative for the
observations to be actually from that particular distribution. We
require a rigorous procedure \cite{shalizi_powerlaw} to test the
validity of our sample following the specified distribution.

After hypothesizing the empirical distribution from the optimal
power law model, we simulate a similar sample from that particular
distribution. The optimal power law model for the simulated sample
is estimated by minimization of the relevant KS statistic over the
values of $x_{min}$. Thereby, we obtain the optimal value for the KS
statistic for this particular sample. If this value is greater than
or equal to the corresponding value obtained from the actual data,
it is an evidence in favour of the real data being from the best
fitted power-law distribution. Otherwise, it is rather unlikely that
the data is actually from the hypothesized power law distribution.

We generate a large number of samples with the same size as that of
the data and calculate the fraction of samples, where the optimal KS
statistic exceeds the one for the real data. We denote this fraction
as the $p$-value of our test statistic. If this $p$-value is large,
say close to one, then evidently the real data is from the best
fitted power law distribution. On the other hand, if this $p$-value
is close to zero, we fail to accept our hypothesis of data being
drawn from a power law distribution. In terms of the level of the
test, if the $p$-value is less than the specified level of the test,
the null hypothesis of the data following a power law distribution
is rejected.

We repeat this entire exercise for the null hypothesis of data
following a Tsallis $q$-exponential distribution as well.



\section{Data Analysis and Results}
\label{section_results}
\subsection*{Data Description}
The Indian Census is conducted once in a decade. We have detailed
data \cite{indiancensus} for the year of 2001. According to the
census conducted on the first day of March, 2001, the population of
India stood at 1,027,015,247 persons. In that census data, there is
a complete enumeration for the population of 4378 Indian urban
agglomerations, 35 of which, have a population greater than a
million. The data shows that only 27.86 percent of Indians live in
these urban agglomerations; the rest of the Indians live in numerous
rural agglomerations.

The People's Republic of China conducted censuses in 1953, 1964, and 1982. In 1987 the government announced that the fourth national census would take place in 1990 and that there would be one every ten years thereafter. The 1982 census, which reported a total population of 1,008,180,738, is generally accepted as significantly more reliable, accurate, and thorough than the previous two. At the 2000 census, the total population stood at approximately  1.29533 billion, which is about 22\% of total population in the world. 36\% of the Chinese population used to reside in urban agglomerations in 2000. We use the data \cite{china_data} from 1990 and 2000 census.

\begin{table}
\begin{tabular}{|c|c|c|c|c|c|c|c|c|}
\hline Study No. & Data-set & n & $x_{min}$ & $x_{max}$ & Mean & Median
& Quartile 1 &
Quartile 3 \\
\hline I & India (2001) & 3307 & 10.00 & 16434.39 & 84.30 & 23.42&
15.39 & 48.23
\\
\hline II& India (2001) & 174 & 203.38 & 16434.39 & 956.38 & 430.50
& 267.66 & 865.55
\\
\hline III& India (1991) & 162 & 160.50 & 12596.24 & 759.45 & 365.31
& 219.75 & 654.49
\\
\hline IV& India (1981) & 152 & 120.42 & 9194.02 & 576.86 & 277.22 &
171.68 & 519.37
\\
\hline V& China (2000) & 1462 & 50.08 & 14230.99 & 298.27 & 136.63 &
80.86 & 265.42
\\
\hline VI & China (2000) & 514 & 200.10 & 14230.99 & 658.95 & 358.81
& 254.67 & 578.36
\\
\hline VII & China (1990) & 1345 & 25.02 & 7821.79 & 156.33 & 68.71
& 44.23 & 128.96
\\
\hline VIII & China (1990) & 280 & 151.58 & 7871.79 & 503.05 &
274.58 & 189.83 & 456.18
\\
\hline
\end{tabular}
\caption{\textbf{Data Description}: Values of x (city-population)
are reported in units of thousands. The left truncation of the data
is determined through the value of $x_{min}$. \\
Source:
\cite{indiancensus} and \cite{china_data}} \label{table_for_data}
\end{table}

In general, the theories for modeling the city-size distribution
does not differentiate between an urban agglomeration and a rural
one. However, the census data does not disclose the complete
enumeration of the sizes for the rural agglomerations. Therefore, we
analyze the size-distribution of the urban agglomerations alone. In
the data, there are many towns with lesser number of inhabitants
compared to that of many villages. If we consider the data for urban
agglomerations in its entirety, this would lead to a biased data set
for the population agglomerates as a whole and hence, to a biased
set of estimates for the studied statistical models. This suggests
us to consider the urban agglomerates over certain minimum value. We
decide to set it at 10,000 for the Indian census data for the year of 2001. There is a trade-off involved in finding
this minimum cut off for a town's population size. The choice of a
rather high value (say 20,000) causes us to loose a large fraction
(nearly half) of our data set; whereas choice of a lower value would
accentuate the problem of biased data-set. Most importantly, a small
movement of the minimum value to either direction would not alter
our estimates even quantitatively.

For the censuses of 1981 and 1991, we do not have the complete
enumeration of the population figures in the Indian cities. However,
we find the individual data for the population of Indian cities with
a certain minimum number of inhabitants. For example, in 1991, we
have individual data  regarding 185 urban agglomeration with the
total population of 125,457,068 persons; while the total urban
population of India in 1991 was 217,611,012. Moreover, individual
figures of all the cities above the population of 160,000 are
included in these data. Therefore, we set the minimum value to
160,000 to left-censor our data-set. Further, to have a comparative
study among the data from 1981, 1991 and 2001, we also work with all
the Indian cities in 2001 with at least 200,000 dwellers.

We also use the individual data \cite{china_data} on the population
of urban agglomerations in China for the years of 1990 and 2000.
We work with two different values for $x_{min}$ in both of these data-sets.
The lower cut-offs (cases V and VII for the years 2000 and 1990) are
chosen such as to include the data-set in its entirety baring a few
outliers. The relatively higher cut-offs (as in cases  VI and VIII for
the years 2000 and 1990) are selected to compare the corresponding figures
for the top ranking cities alone. It may be mentioned here that the distributions of Tsallis
$q$-exponential and Pareto differ only at the lower level. We
tabulate the descriptive statistics regarding all the data-sets in
Table \ref{table_for_data} for all the eight cases considered.

\subsection*{Results}

We report the estimates corresponding to all the eight different
cases studied in this paper in Table \ref{table_for_result}. We have
elaborated the estimation procedure in Section \ref{methodology}.
The usual linear fit estimate using the Pareto Distribution is given
by $\widehat{\alpha}_{lf}$; whereas the maximum likelihood estimate
of the same is denoted as $\widehat{\alpha}_{MLE}$. It is emphasized here that the linear fit estimate, $\widehat{\alpha}_{lf}$, is not suitable for a proper estimation of the scaling parameter. As most of the prevalent literature has based their conclusion based on this measure, we compute it merely to measure the quantity and direction of the bias in this estimate. We find a
considerable bias in the linear fit estimate compared to the
corresponding one obtained using the technique of MLE. Also, the MLE
of the parameters for the Tsallis $q$-exponential distribution are
expressed as $\widehat{\theta}_{MLE}$ and  $\widehat{\sigma}_{MLE}$
in Table \ref{table_for_result}.  For each study, we plot the
corresponding data-set and the fitted CCDFs. The graphical
representation shows that in all the cases studied the estimated
Pareto distribution and the fitted Tsallis $q$-exponential
distribution are almost identical. Therefore, we restrict our
attention to Pareto distribution alone for further investigation.

\begin{table}[h]
  \begin{tabular}{|c|c|c|c|c|c|c|c|c|}
    \hline
 & \multicolumn{2}{|c|}{Pareto}  & \multicolumn{2}{|c|}{Tsallis}
   & \multicolumn{4}{|c|}{Minimized KS Distance Estimate with}\\
\cline{6-9}   Study No. & \multicolumn{2}{|c|}{Distribution} &
\multicolumn{2}{|c|}{Distribution}& \multicolumn{2}{|c|}{Pareto
Dist.} &
           \multicolumn{2}{|c|}{Tsallis Dist.} \\
           \cline{2-9}
           &  $\widehat{\alpha}_{LF}$ &
$\widehat{\alpha}_{MLE}$ & $\widehat{\theta}_{MLE}$ &
    $\widehat{\sigma}_{MLE}$ & statistic & $p$-value & statistic & $p$-value\\
\hline I & 1.9923 & 1.8827 & 0.8827 & 0.0084 &  0.0348 & 0.000 &
0.0348 & 0.000
\\       & (0.0010)& (0.0153)& (0.0002)& (0.0132)& & & &
\\
\hline II & 1.9133 & 2.0320 & 1.0320& 0.0530 & 0.0420 & 0.290 &
0.0420 & 0.286
\\
          & (0.0073) & (0.0782)& (0.0164) & (0.0709)& & & &
\\
\hline III& 1.8946 & 2.0601 & 1.0601 & 0.0192 & 0.0648 & 0.005 &
0.0648 & 0.006
\\
          & (0.0083)& (0.0838) & (0.0025) & (0.0692)& & & &
\\
\hline IV & 1.8893 & 1.9909 & 0.9909 & 0.0224 & 0.0557 & 0.052 &
0.0557 & 0.053
\\ & (0.0085) & (0.0804)& (0.0027) & (0.0675) & & & &
\\
\hline V & 1.8976 & 1.8480 & 0.8480 & 0.0120 & 0.0568 & 0.000 &
0.0568 & 0.000
\\
         & (0.0036)& (0.0222)& (0.0005) & (0.0167)& & & &
         \\
\hline VI &  1.7544 & 2.2975 & 1.2975 & -0.1056 & 0.0217 & 0.531 &
0.0217 & 0.871
\\
          & (0.0018) & (0.0572) & (0.0424) & (0.0550)  &  & &  &
\\
\hline VII & 1.8967 & 1.8241 & 0.8241 & 0.0076 & 0.0682 & 0.000 &
0.0682 & 0.000
\\
          & (0.0031)& (0.0225) & (0.0001) & (0.0167) &  & &  &
\\
\hline VIII & 1.7701 & 2.2308 &  1.2308 & 0.0666 & 0.0229 & 0.913 &
0.0229 & 0.913
\\
            & (0.0032) & (0.0736) & (0.0299)& (0.0675) & & & &
\\
    \hline
  \end{tabular}
  \caption{Estimates from various data-sets considered: The standard errors are in parenthesis}
  \label{table_for_result}
\end{table}

In case of India, the estimated exponent ($\widehat{\alpha}_{MLE}$)
is within the range of $[1.8827, 2.0601]$ for different cases
considered. The value is a good approximation of the theoretical
predicted value of $2$ for $\alpha$. For China,
$\widehat{\alpha}_{MLE}$ depends on the chosen value of $x_{min}$.
It is in the range of $[1.8241, 2.2975]$ contingent on the choice of
$x_{min}$. For higher values of $x_{min}$, the estimate is bigger.
It might be interesting to compare it with other studies. In case of
cities of Brazil \cite{pro1} with $x_{min}$ as 30,000 the estimated
value of $\alpha$ is found to be 2.41 for 1970 and 2.36 for
1980-2000. The corresponding linear fit estimate for 2400 U.S.
cities \cite{switzerland} is 2.1 for the year of 2000 and that for
Switzerland stands at 2.0. The estimates using the data from Japan
\cite{japan} is rather interesting. It is shown that the Zipf's law
($\alpha = 2$) holds for the period of 1970-2000. Before and after
this time period, $\alpha$ is significantly greater than 2. Using
the KS statistic, we have computed the optimal estimate in the class
of best fitted models and the endogenous value for $x_{min}$. In all
the cases, it is almost same as the exogenously fixed value of
$x_{min}$ in the data.

However, the data-set may contain a lot of non-sampling errors. The
scatter plot reveals that the percentage of variation in our
estimates is quite large, if we consider the slope of the fitted
line neglecting a small number of observations. This is indicative
of a poor quality of the data. So, we consider only two digits after
the decimal place for all our estimates as significant. The
interpretations for all the computed estimates in Table
\ref{table_for_result} should be modified accordingly.

 A natural question may arise whether there is any other
suggested distribution from the exponential family that explains the
data better. There is an indication \cite{weibl} that the Weibull
distribution provides us a satisfactory adjustment for some ranges
of the data. We carry out a likelihood maximization (LM) test
\cite{crrao} with our data-set, where the null hypothesis of data
being from the Pareto distribution is tested against the alternative
of various other statistical distributions, namely Weibull,
Exponential, Exponential with a cutoff and log-normal. The $p$-value
of the test statistic is always equal to one, which indicates that
the null hypothesis is strongly accepted against the specified
various alternatives. However, there is a word of caveat regarding
this observation. The result does not imply that the assumption of
the data following the Pareto distribution is justified. It is only
a better description of the data over the other specified
alternatives.

To test the validity of Pareto distribution,  we compute the KS
statistic as the distance between the empirical CDF and that of a
fitted distribution as elaborated in Section \ref{methodology}. The
relevant $p$-values in Table \ref{table_for_result} reveals that
when we analyse the data comprising the sample of Indian cities with
a higher $x_{min}$, the null hypothesis of Pareto distribution is
accepted at the 5\% critical level for all the years. But the null
hypothesis can not be accepted if we consider the full sample for
the year of 2001. In case of Chinese cities, the null hypothesis is
again rejected with the full sample. However, it is well-accepted at
any critical level, if the short sample with higher ranked cities is
considered.

Finally, the graphical representation discloses that for both the
countries of India and China, the comparatively higher ranked cities
have disproportionately more dwellers compared to rather lower
ranked ones. In general, the theories for size distribution implying Zipf's law
does not take into account any rural to urban migration, while modeling the distribution of urban agglomerations. However, it is an important phenomenon in the developing countries like India and China. In these countries,  various Economic opportunities drive \cite{harris} people from rural agglomerations to urban ones and also from smaller towns to larger cities. A theory, taking into account this factor, can explain the size distribution of the entire sample for the Indian (or Chinese) urban agglomerations.

\section{Conclusion}

In this work, we have shown that the city-size distribution for both
the countries of India and China follow the Zipf's law, if only we
work with a more trimmed sample keeping the $x_{min}$ quite high. We
have estimated the scaling exponent, by the linear fit method as
well as by a more accurate technique of Maximum Likelihood
Estimator, which is found to be nearly 2 as predicted. The maximum
likelihood estimation with Tsallis $q$-exponential distribution is
also performed, although the estimated CDF of this distribution is
identical with its Pareto counterpart. The novelty of our work lies
in the goodness-of-fit tests. The Kolmogorov-Smirnov statistic for
the sample with the computed $p$-value implies that the full sample
does not follow a Pareto or $q$-exponential Tsallis distribution too
well. However, it gives a good approximation for a restricted sample
with top ranking cities.

{\it Acknowledgement: The authors thank Soumyasree Bandyapadhyay for compilation of data for this project.}


\newpage

\begin{figure}[ht]
\centering \subfigure[Indian Cities with population  over 10,000 in
2001]{
\includegraphics[scale=0.40]{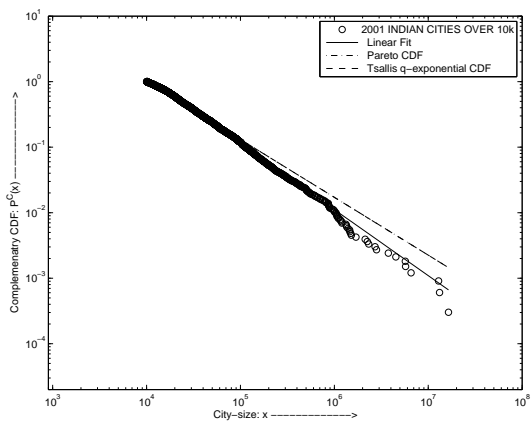}
\label{fig:subfig11} } \subfigure[Indian Cities with population
 over 200,000 in 2001]{
\includegraphics[scale=0.40]{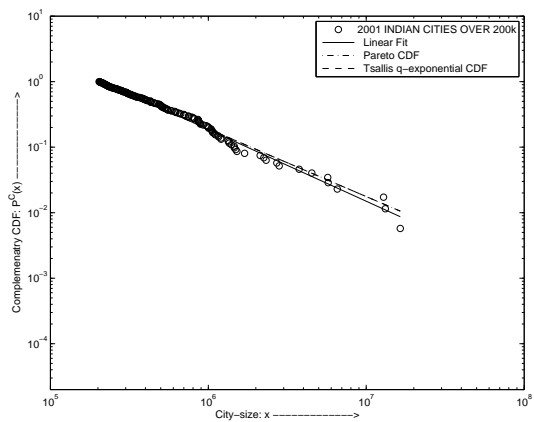}
\label{fig:subfig12} } \subfigure[Indian Cities with population over 160,000 in 1991]{
\includegraphics[scale=0.40]{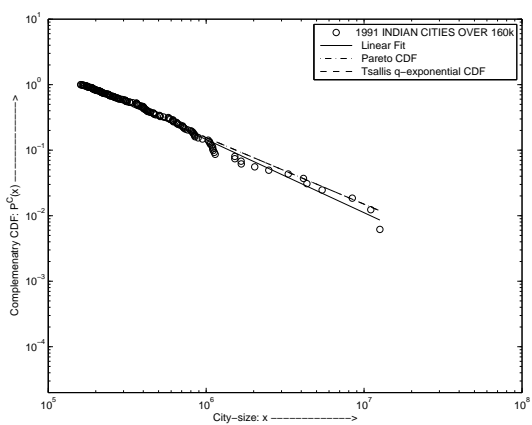}
\label{fig:subfig13} } \subfigure[Indian Cities with population over 120,000 in 1981]{
\includegraphics[scale=0.40]{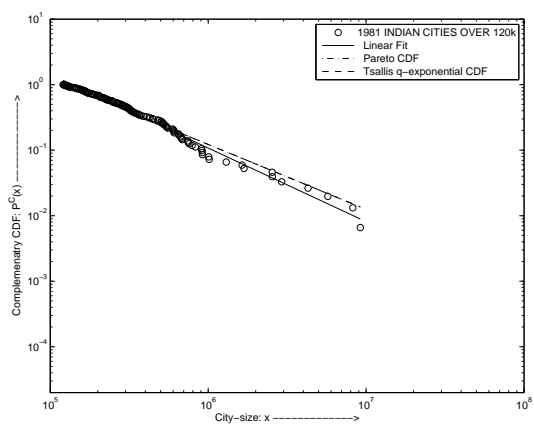}
\label{fig:subfig14} } \label{fig:subfigureExample1}
\caption[]{Indian Cities: 1981-2001}
\end{figure}

\begin{figure}[ht]
\centering \subfigure[Chinese Cities  with population  over 50,000 in 2000]{
\includegraphics[scale=0.40]{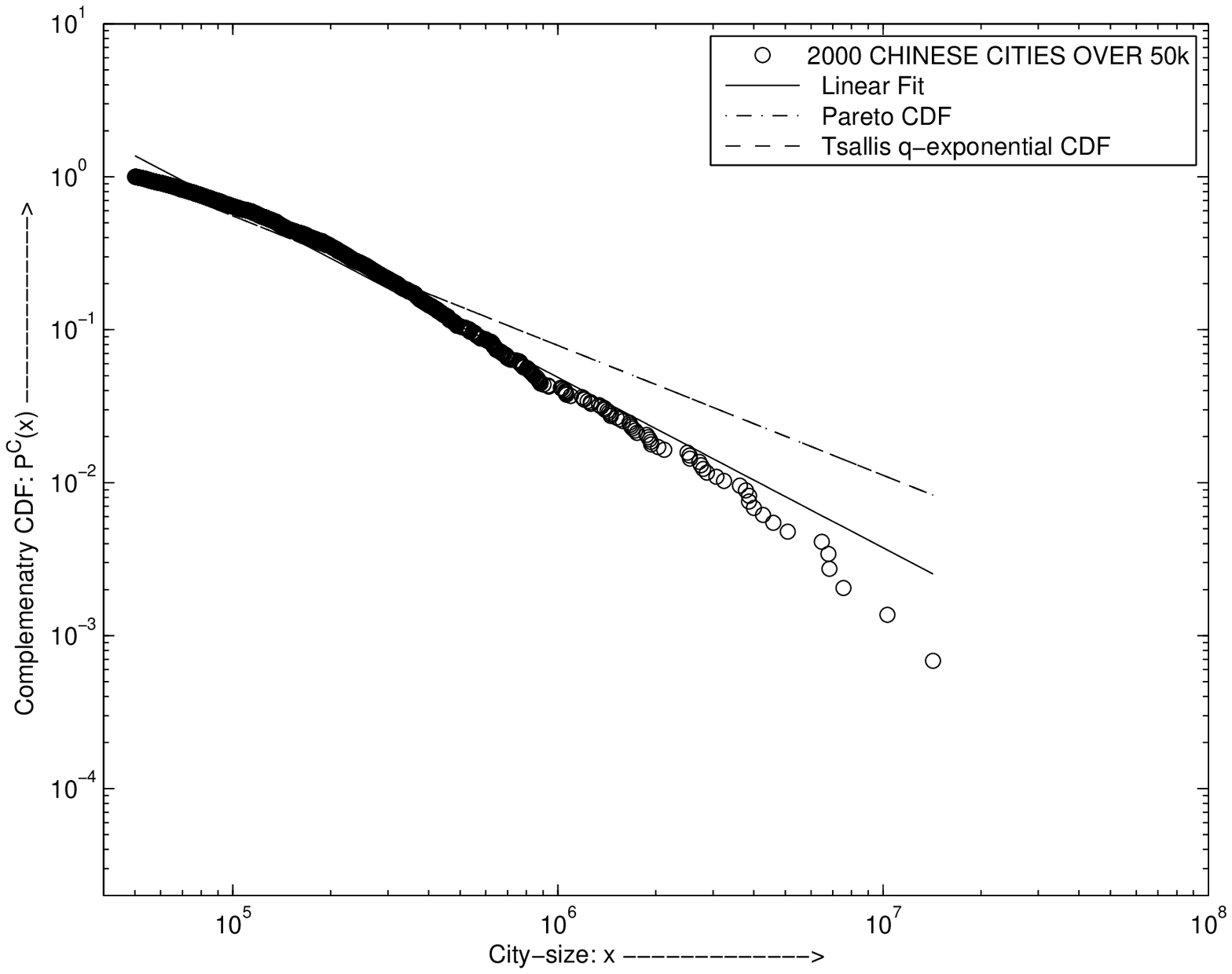}
\label{fig:subfig21} } \subfigure[Chinese Cities with population
over 200,000 in 2000]{
\includegraphics[scale=0.40]{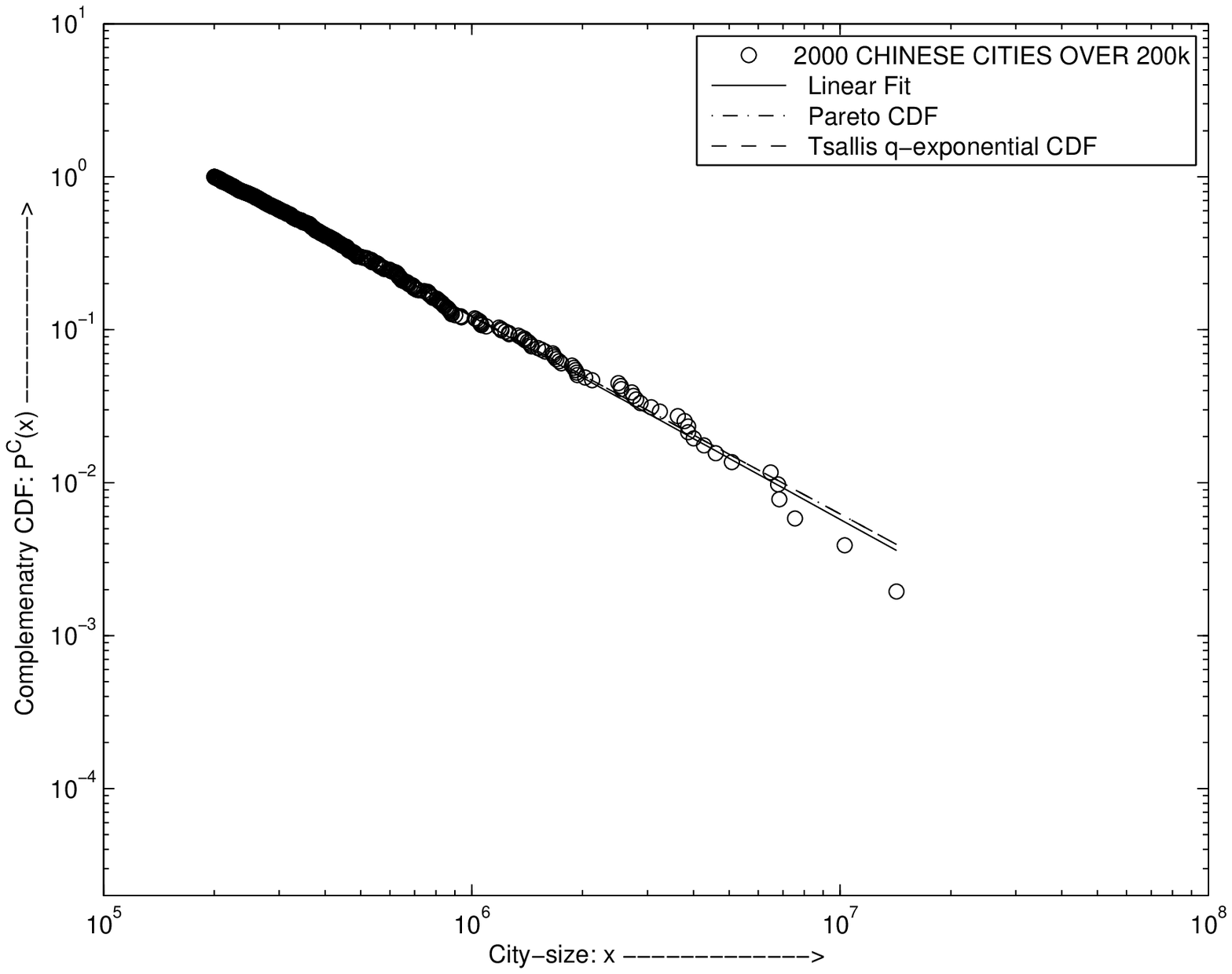}
\label{fig:subfig22} } \subfigure[Chinese Cities with population
 over 25,000 in 1990]{
\includegraphics[scale=0.40]{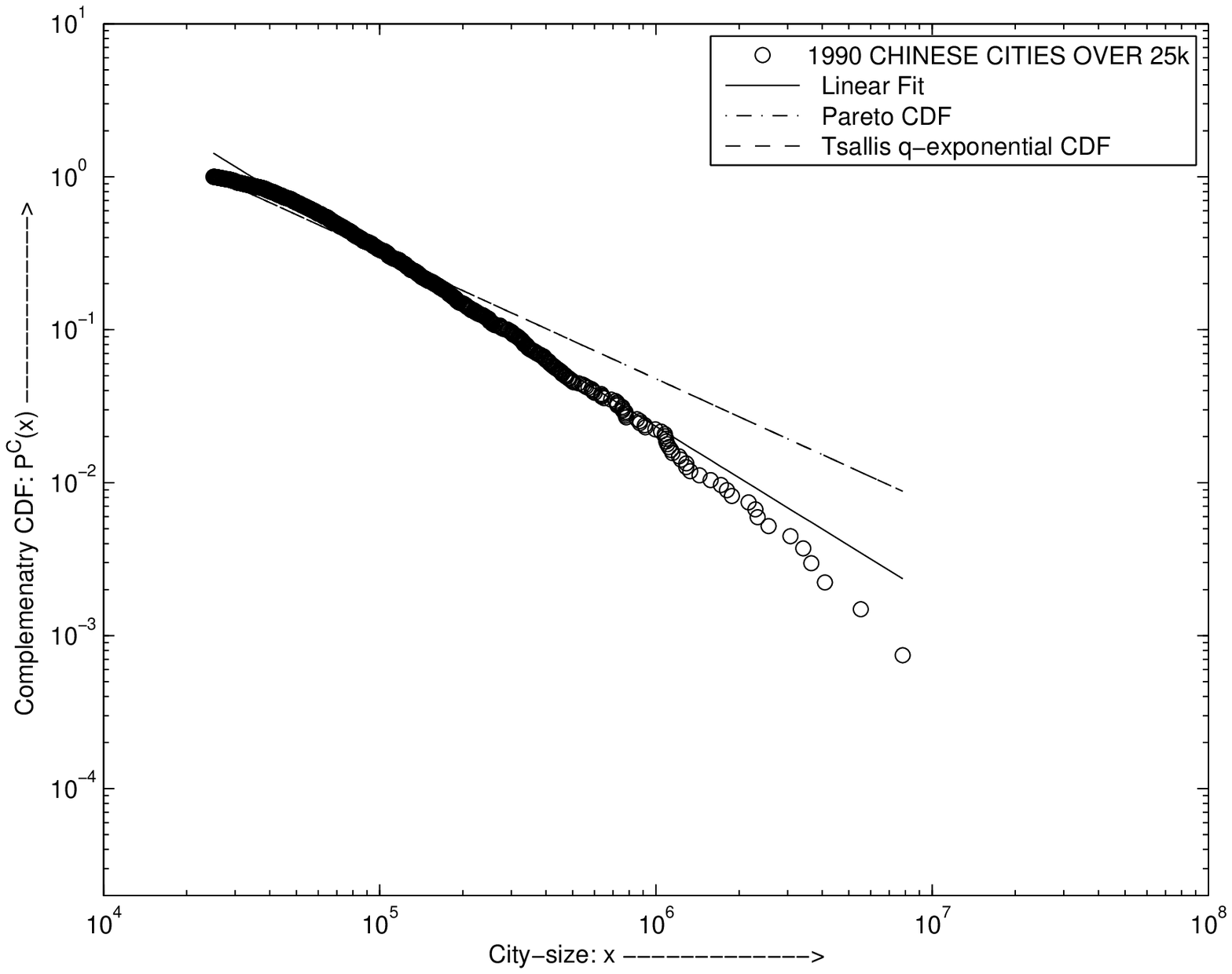}
\label{fig:subfig23} } \subfigure[Chinese Cities with population
 over 150,000 in 1990]{
\includegraphics[scale=0.40]{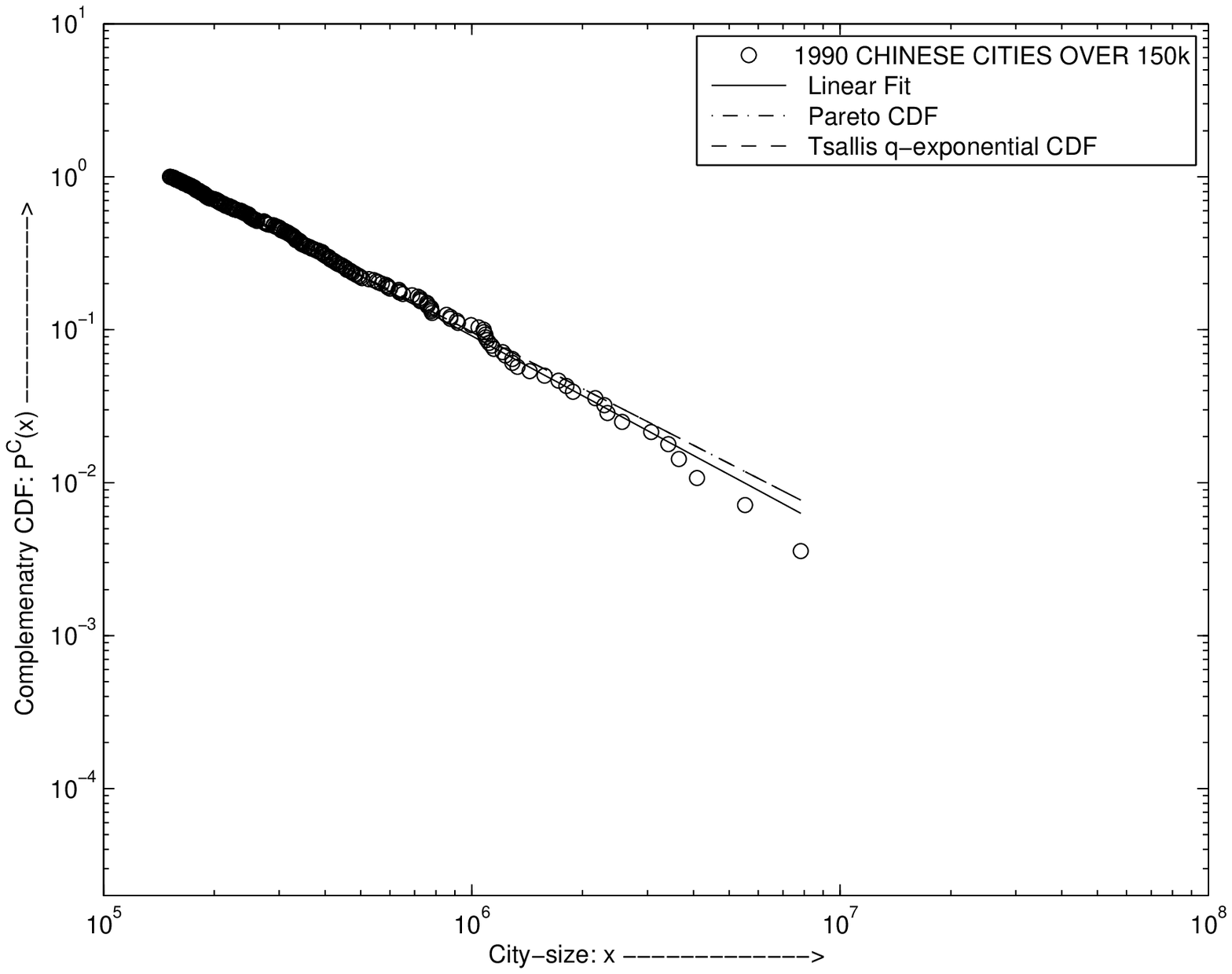}
\label{fig:subfig24} }
 \label{fig:subfigureExample2} \caption[Optional
caption for list of figures]{Chinese Cities: 1990 and 2000}
\end{figure}

\end{document}